\documentclass[usenatbib]{mn2e}

\newcommand{\msun}{\mbox{M$_\odot$}}
\usepackage{graphicx}
\usepackage{amssymb}

\title[A 0535+26]{A 0535+26 : back in business}

\author[M.J. Coe et al.]{M. J.~Coe$^{1}$, P. ~Reig$^{2}$,
V.A. McBride$^{1}$, J.L. Galache$^{1}$ \& J. Fabregat$^{3}$  \\
$^{1}$School of Physics and Astronomy, Southampton University, SO17 
1BJ, UK\\
$^{2}$ IESL (FORTH) \& Physics Department, University of Crete, 71003, Heraklion, Greece\\
$^{3}$ Observatori Astronomic, Universidad de Valencia, P.O. BOX 22085, E46071 Valencia, Spain}
\begin{document}

\date{Feb 2006}

\pagerange{\pageref{firstpage}--\pageref{lastpage}} \pubyear{2002}

\maketitle

\label{firstpage}

\begin{abstract}

In May/June 2005, after 10 years of inactivity, the Be/X-ray binary
system A 0535+26 underwent a major X-ray outburst. In this paper data are
presented from 10 years of optical, IR and X-ray monitoring showing
the behaviour of the system during the quiescent epoch and the lead up
to the new outburst. The results show the system going through a
period when the Be star in the system had a minimal circumstellar disk
and then a dramatic disk recovery leading, presumably, to the latest
flare up of X-ray emission. The data are interpreted in terms of the
state of the disk and its interaction with the neutron star companion.

\end{abstract}


\section{Introduction and background}

The Be/X-ray systems represent the largest sub-class of massive X-ray
binaries.  A survey of the literature reveals that of the 115
identified massive X-ray binary pulsar systems (identified here means
exhibiting a coherent X-ray pulse period), most of the systems fall
within this Be counterpart class of binary.  The orbit of the Be star
and the compact object, presumably a neutron star, is generally wide
and eccentric.  X-ray outbursts are normally associated with the
passage of the neutron star close to the circumstellar disk (Okazaki
\& Negueruela 2001). A review of such systems may
be found in Coe et al. (2000).

The source that is the subject of this paper, A 0535+26, was
discovered by Ariel V during a Type II outburst (see Stella, White \&
Rosner 1986 for a description of different outburst types) in 1975 (Coe et
al. 1975, Rosenberg et al. 1975).  The primary is an O9.7IIIe star (Steele
et al. 1998) and its optical and infrared (IR) emission has been the subject
of many papers in an attempt to decode the behavioural patterns of this system.
Extensive pieces of work by Clark et al. (1998) and Haigh et
al. (2004) using both photometric and spectroscopic data in the UV-IR
range have presented data sets showing assorted levels of
variation within the optical star. In contrast, throughout the last decade
(1995-2004) there have been no reported detections of the
source at X-ray wavelengths. However, in May/June 2005
the source was detected undergoing significant X-ray
activity by the SWIFT (Tueller et al. 2005) and RHESSI (Smith et al.
2005) observatories. In this paper we report on the recent
developments in the optical and IR fluxes in the context of the
re-emergence of A 0535+26 as an active X-ray source.

\section{Optical data}

H$\alpha$ data have been collected over the last 10 years from a
series of telescopes. Some of the early data have already been
published in Haigh et al. (2004) but are presented here again
to provide the context for the changes that have subsequently taken
place in the source. 

The dates and properties of the H$\alpha$ line are presented in
Table~\ref{tab:obs}. In this table the following telescopes and
configurations have been used:

\begin{itemize}

\item JKT - 1.0m telescope, La Palma observatory (Spain), RBS
spectrograph, TEK4 detector, R1200Y grating
\item INT - 2.5m telescope, La Palma observatory (Spain), IDS instrument, EEV5 detector, R1200R grating
\item Ski - 1.3m telescope, Mt. Skinakas Observatory (Crete),
spectrograph, SITe detector, 1301 l/mm grating
\item SAAO - 1.9m telescope, Sutherland Observatory (South Africa), spectrograph, SITe detector, 1200 l/mm grating

\end{itemize}

\begin{table}
\begin{center}
\caption{Table of H$\alpha$ measurements. See text for details of
observatory/instrument used.}
\begin{tabular}{ccccc}
\hline
Obs. date & Telescope& H$\alpha$EW& V/R& Peak \\
&&(\AA)&ratio&Separation \\
&&&& (km/s) \\
\hline
28 Feb 1996&JKT &-8.2$\pm$0.2&0.76&247$\pm$5 \\
28 Oct 1997&JKT &-7.3$\pm$0.1&1.09&251$\pm$5 \\
26 Nov 1998&INT &-0.9$\pm$0.1&1.59&471$\pm$5 \\
24 Apr 1999&INT &-0.5$\pm$0.1&1.33&429$\pm$5 \\
15 Aug 2000&INT &-5.1$\pm$0.1&1.06&256$\pm$5 \\
16 Oct 2000&Ski &-8.6$\pm$0.2&-& - \\
12 Sep 2001&Ski &-9.2$\pm$0.2&-& - \\
08 Oct 2001&Ski &-10.5$\pm$0.3&-& - \\
11 Nov 2001&SAAO&-9.0$\pm$0.3&1.17&210$\pm$5 \\
11 Sep 2002&Ski &-13.1$\pm$0.4&-& - \\
13 Dec 2002&SAAO&-11.6$\pm$0.2&1.30&214$\pm$5 \\
06 Oct 2003&Ski &-11.0$\pm$0.5&-& - \\
24 Oct 2004&Ski &-13.1$\pm$0.4&-& - \\
21 Mar 2005&SAAO&-12.8$\pm$0.2&1.34&179$\pm$5 \\
16 Aug 2005&Ski &-14.7$\pm$0.3&-& - \\
17 Aug 2005&Ski &-14.9$\pm$0.2&-& - \\
\hline
\end{tabular}
\label{tab:obs}
\end{center}
\end{table}

\begin{figure}\begin{center}
\includegraphics[width=85mm,angle=-0]{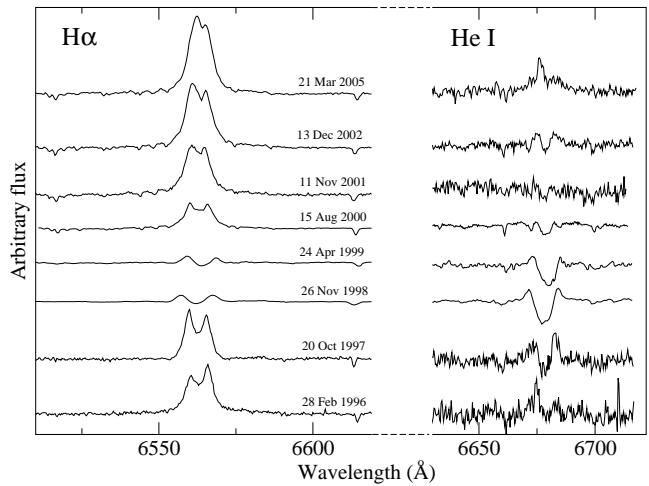}
\caption{The evolution of the H$\alpha$  and 6678\AA ~HeI profiles over 10 years. The
feature at 6614\AA ~is a DIB.}
\label{fig:ha}
\end{center}
\end{figure}

All of the H$\alpha$ lines showing resolved structures are presented
in Figure~\ref{fig:ha}. They are arranged chronologically in this
figure, with the height of the H$\alpha$ line linearily related to its
equivalent width. A clear evolution of the Balmer line over the 10 year
period is seen, progressing from modest emission, through an almost
complete disk loss phase, back to strong emission. Comparison with the
previous 10 years worth of data presented in Clark et al. (1998) shows
the current emission line strength to be comparable with the largest
values seen at any time in the last two decades. 
In addition to the strength of the H$\alpha$ equivalent width, the V/R
ratio was determined where possible. These values are also included in
Table~\ref{tab:obs}. 

It is worth noting that the He 6678\AA ~line
shows a similar pattern of behaviour to the H$\alpha$ line -- going
through an almost complete loss of emission in 1998, and then slowly
reverting to a clear emission feature by 2005.

\section{IR data}

\begin{figure}\begin{center}
\includegraphics[width=75mm,angle=-0]{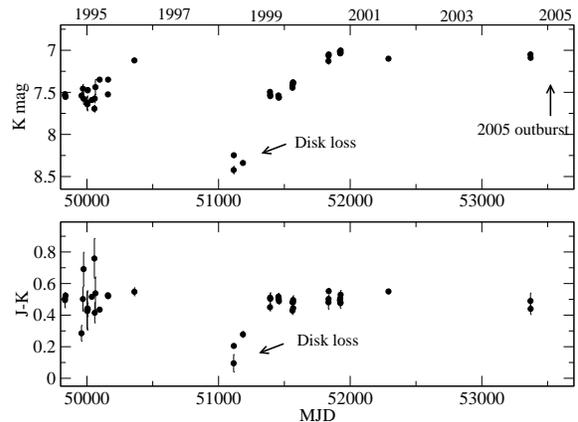}
\caption{The IR K band flux (upper panel) and colour (J-K) (lower
panel) over the 10 years 1995 -- 2005.}
\label{fig:ir}
\end{center}
\end{figure}

IR data on A 0535+26 have been collected since 1987 from the 1.5m
Telescopio Carlos Sanchez in Tenerife, Spain. The data were reduced
following the procedure described by Manfroid (1993). Instrumental
values were transformed to the TCS standard system (Alonso, Arribas \&
Martinez-Roger 1998). The dates and results of the
IR observations are presented in Table~\ref{tab:ir}. and shown in 
Figure~\ref{fig:ir}.

\begin{table*}
\begin{center}
\caption{Table of IR measurements taken from the Telescopio Carlos
Sanchez in Tenerife.}

\begin{tabular}{cccccccc}
\hline
Date & MJD & J & J error& H & H error& K & K error \\
\hline
02-01-95 &   9720.663 &   7.994 & 0.005 & 7.756 & 0.006 & 7.479 & 0.004 \\
03-01-95 &   9721.54 & 7.998 & 0.007 & 7.75 &  0.001 & 7.504 & 0.001 \\
04-01-95 &  9722.625 &   8.035 & 0.005 & 7.786 & 0.005 & 7.522 & 0.003 \\
05-01-95 &  9723.524 &   8.030 &  0.014 & 7.785 & 0.013 & 7.527 & 0.008 \\
25-04-95 &  9833.355 &   8.029 & 0.002 & 7.758 & 0.003 & 7.522 & 0.002 \\
29-04-95 &  9837.363 &   8.078 & 0.005 & 7.828 & 0.001 & 7.554 & 0.009 \\
11-10-95 &  10002.69 &   8.070 &  0.004 & 7.758 & 0.003 & 7.63 &  0.003 \\
11-10-95 &  10002.72 &   8.074 & 0.003 & 7.746 & 0.002 & 7.641 & 0.003 \\
14-10-95 &  10005.73 &   7.917 & 0.002 & 7.702 & 0.004 & 7.475 & 0.004 \\
13-01-96 &  10096.59 &   7.784 & 0.005 & 7.562 & 0.005 & 7.349 & 0.001 \\
16-03-96 &  10159.44 &   8.050 &  0.009 & 7.695 & 0.006 & 7.524 & 0.006 \\
17-03-96 &  10160.45 &   7.869 & 0.004 & 7.603 & 0.003 & 7.349 & 0.003 \\
27-10-98 &  11114.652 &  8.518 & 0.033 & 8.469 & 0.016 & 8.423 & 0.044 \\
28-10-98 &  11115.709 &  8.452 & 0.008 & 8.316 & 0.004 & 8.247 & 0.000 \\
 30-7-99 &  11390.734 &  7.944 & 0.014 & 7.710 & 0.016 & 7.494 & 0.019 \\
 31-7-99 &  11391.731 & 8.038 & 0.018 & 7.764 & 0.017 & 7.527 & 0.017 \\
 1-8-99  & 11392.732 & 8.048 & 0.033 & 7.798 & 0.026 & 7.545 & 0.020 \\
 02-10-99 & 11454.616 & 8.055 & 0.015 & 7.794 & 0.015 & 7.538 & 0.015 \\
 03-10-99 & 11455.707 & 8.046 & 0.014 & 7.790 & 0.015 & 7.541 & 0.015 \\
 04-10-99 & 11456.703 & 8.066 & 0.017 & 7.822 & 0.011 & 7.562 & 0.011 \\
 06-10-99 & 11458.747 & 8.048 & 0.016 & 7.804 & 0.010 & 7.559 & 0.011 \\
 17-01-00 & 11561.566 & 7.874 & 0.023 & 7.642 & 0.014 & 7.445 & 0.010 \\
 17-01-00 & 11561.578 & 7.898 & 0.024 & 7.641 & 0.015 & 7.417 & 0.012 \\
 21-01-00 & 11565.592 & 7.863 & 0.015 & 7.601 & 0.016 & 7.381 & 0.019 \\
 22-01-00 & 11566.512 & 7.832 & 0.023 & 7.596 & 0.014 & 7.388 & 0.015 \\
 22-01-00 & 11566.632 & 7.884 & 0.023 & 7.649 & 0.015 & 7.388 & 0.014 \\
 16-10-00 &  11834.772 &  7.609 & 0.023 & 7.378 & 0.032 & 7.128 &
0.037 \\
 17-10-00 &  11835.760 &  7.567 & 0.015 & 7.324 & 0.010 & 7.065 &
0.011 \\
 18-10-00 &  11836.700 &  7.602 & 0.014 & 7.318 & 0.010 & 7.050 & 0.010\\
 11-01-01 &  11921.479 &  7.512 & 0.032 & 7.273 & 0.014 & 7.017 & 0.015\\
 14-01-01 &  11924.506 &  7.544 & 0.023 & 7.282 & 0.023 & 7.037 & 0.023\\
 15-01-01 &  11925.511 &  7.500 & 0.024 & 7.259 & 0.023 & 7.024 & 0.023\\
 16-01-01 &  11926.506 &  7.513 & 0.010 & 7.278 & 0.010 & 7.033 & 0.015\\
 17-01-01 &  11927.503 &  7.533 & 0.023 & 7.259 & 0.014 & 7.003 & 0.014\\
 15-01-02 &  12290.568 &  7.65 & 0.01 & 7.44 & 0.01 & 7.10 & 0.01\\
 29-12-04 &  13369.651 &  7.54 & 0.04 & 7.28 & 0.03 & 7.05 & 0.03\\
 30-12-04 &  13370.616 &  7.53 & 0.03 & 7.27 & 0.02 & 7.09 & 0.02\\

\hline
\end{tabular}

\label{tab:ir}
\end{center}
\end{table*}

\section{X-ray data}

X-ray monitoring data were collected by the All Sky Monitor (ASM)
instrument on the Rossi X-ray Timing Explorer (RXTE) observatory. The
data consist of daily flux averages covering the X-ray energy range
1.3 -- 12.1 keV. The measurements are interrupted once per year for a
few days while the sun passes within a few degrees of A 0535+26. There
is no evidence in the daily data averages for any significant detection of the source.



\begin{figure}\begin{center}
\includegraphics[width=55mm,angle=90]{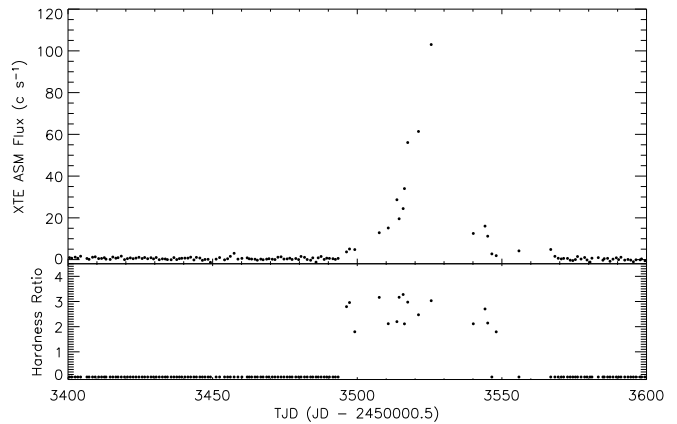}
\caption{RXTE/ASM flux (upper panel) and colours 
(HR2, lower panel) during the outburst.}
\label{fig:out3}
\end{center}
\end{figure}

ASM data from just the outburst period are shown in
Figure~\ref{fig:out3}. In this figure the vertical flux scale in the
top panel has been presented showing a clear
exponential rise with a doubling timescale of $\sim$7d. Though flux
measurements during the peak to decline phase are much fewer, it
is still possible to estimate the decline timescale. This was much
faster than the rise, with the flux falling by a factor of 2 in no
more than
$\sim$2d. The total duration of the outburst was $\sim$70d. In the
lower panel the ratio of two separate X-ray bands (called the HR2
ratio: (5 -- 12 keV)/(3 -- 5 keV)) is presented, showing little evidence for
any hardness changes throughout the outburst.

\begin{figure}\begin{center}
\includegraphics[width=55mm,angle=90]{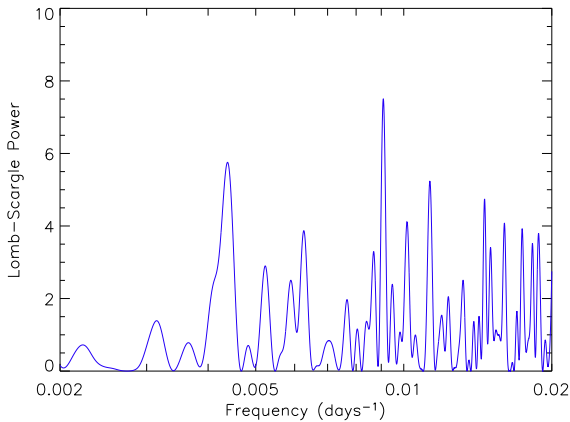}
\caption{Lomb-Scargle power spectrum over the period range 50 -- 500d of all
RXTE/ASM data up to, but not including the May/June 2005 outburst. The power due
to an annual modulation has been removed (see text). The highest peak is at
110.0d.}
\label{fig:ls}
\end{center}
\end{figure}

The $\sim$10 years worth of data, not including the present outburst, were
searched for any very low-level repetitive modulation around the known orbital
period of $\sim$110d (Finger et al. 1994, 1996). A simple Lomb-Scargle analysis of the
raw data for the period MJD = 50087 - 53495 immediately revealed a clear peak at a period around 110d (See
Figure~\ref{fig:ls}).

Various checks were carried out to establish confidence in this 110d
period: 

\begin{itemize}

\item Individual daily averages containing less than 5
measurements (``dwells'') were rejected from the data sample. This
reduces the number of outlying points in the
data. However, this did not significantly affect the position or
strength of the peak in the power spectrum.

\item The data values were randomised within the
exisiting temporal structure to see if the annual gaps were in any way
producing a window function with power in the region of interest.
After 1000 trials the average power spectrum showed no significant
features at, or close to 110d. 

\item When the effects due to the annual modulation were removed from
the lightcurve, the resulting power spectrum (Figure~\ref{fig:ls})
shows a clear peak at 110.0d. The significance of this peak falling
within $\pm$1d of the previously known 110d period is estimated to be
99.94\%.  The significance of a given peak is given by 
$sig$ = 100$\times$$(1-e^{-Z})^M$ (Scargle 1982), where $Z$ is the normalised
power of the peak and $M$ is the number of independent frequencies
within the search limits.

In addition to the Lomb-Scargle periodogram we also search for
periodicities by performing a discrete Fourier transform.  We
extracted a 1-day average light curve from the definite 1-dwell ASM
light curve. The rebinned light curve contains gaps due to the source
being too close to the Sun, rejections of bins with less than 5 data
points, detector failure or other instrumental causes. We filled in
these gaps by interpolation.  However, to prevent the interpolation
procedure from affecting the final power spectrum we allowed
interpolation only when the gap contained less than 1\% of the total
number of points in the light curve. The final light curve then
contained continuous stretched of data separated by large gaps
(typically of several tens of days) cause by the proximity of the
source to the Sun. In order to extract the maximum amount of
information from the light curves we used the discrete (slow) Fourier
transform on each observational segment of data. Since the data
segments had different lengths, the resulting power density spectra
covered different frequency ranges. Due to the typical duration of the
individual segments ($\sim 300$ days. i.e. three cycles in the best
cases) the detection of the 110-d periodicity is weak. Its first
harmonic, however, is more significant. The average of local maxima at
around the expected frequency of the first harmonic results in a peak
that corresponds to 53$\pm$2 days.

\item Finally we note that the orbital
precession period of the RXTE spacecraft is in the range 50 -- 51d
(gradually changing over the 10 years) and hence cannot be responsible
for a peak around 110d.

From Figure~\ref{fig:ls} a clear peak at 110.0$\pm$0.5d is
evident. The period error is determined using the method of Horne \&
Baliunas (1986) in section II.c.

This is consistent with the binary period of 110.3$\pm$0.3d reported
by Finger et al. (1996) from analysis of earlier BATSE data. If the
data are then folded at this period of 110.0d the resulting modulation
is shown in Figure~\ref{fig:fnp}, which shows an apparent low
amplitude sinusoidal modulation with a semi-amplitude of
($0.085\pm0.008$) counts.

Alternatively see Figure~\ref{fig:jose}. In this version of the figure
the folded light curve is obtained from $m$ sets of $n$-binned folded
light curves. To begin with, the light curve is folded at the desired
period and the time values converted into orbital phase values. This
``raw'' folded light curve is then binned into $n$ bins (of size
$1/n$) in the standard way, with each bin starting at phase $a/n$
(with $a = 0,1,2,...n-1$); the error on each bin is taken to be the
standard deviation of the flux values within it. This step is repeated
again, but this time the bins start at phase $a/n + 1/(n \times
m)$. In this way we create $m$ folded light curves, each consisting of
$n$ bins, but differing in the starting phase of their bins, the
general expression of which is $a/n + m/(n \times m)$. Now each bin in
the folded light curves is sub-divided into $m$ bins, all of which
have the same flux value.  The final folded light curve will have $l =
n \times m$ bins, each of which is the average of the bins (at the
same phase) from the $m$ sets of previous lightcurves. The error for
each bin is the standard error of the $m$ values averaged for each
$l$-bin. This method is explained in Laycock (2002)

This method provides an efficient way of generating folded light
curves from poorly sampled data, or data with low S/N. Their shape
will not depend on the starting point at which they are folded and,
although only every $m^{th}$ bin will be independent, spurious flux
values within bins will be evened out creating a smoother
profile. Explanations of the negative effects on pulsar profiles
associated with folding at inapropriate starting points can be found
in Carstairs (1992), while the effects of varying bin widths (albeit
for slightly different purposes) is studied by De Jager et al. (1989).

\begin{figure}\begin{center}
\includegraphics[width=55mm,angle=-90]{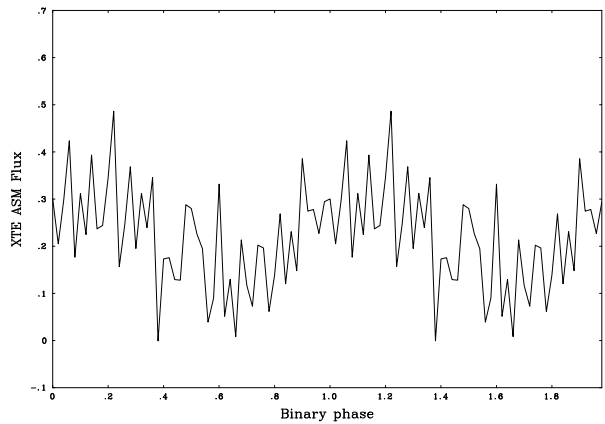}
\caption{RXTE/ASM data (not including the outburst) folded at the period
of 110.0d}
\label{fig:fnp}
\end{center}
\end{figure}

\begin{figure}\begin{center}
\includegraphics[width=55mm,angle=90]{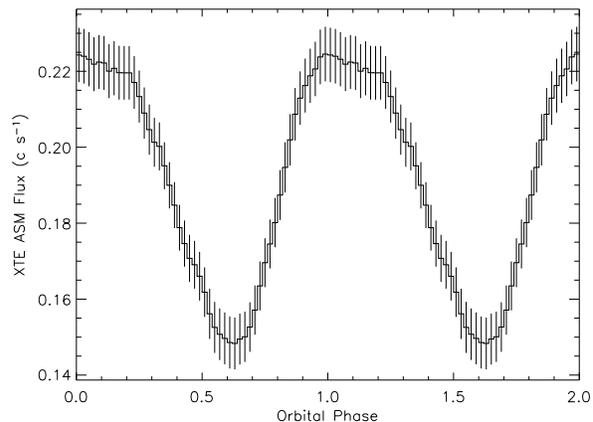}
\caption{RXTE/ASM data (not including the outburst) folded at the period
of 110.0d. The folding technique used here differs from that in
Figure~\ref{fig:fnp} in that the pulse shape of the folded light curve is independent of the point at which one
starts the fold - see text for details.}
\label{fig:jose}
\end{center}
\end{figure}

If this modulation indicates increased X-ray activity associated with periastron
passages of the neutron star, then the ephemeris derived from it is:

\begin{center}
$P_\mathrm{orb} = 110.0\pm0.5$ d\\
$\tau_\mathrm{periastron} = 2450094\pm1$ JD. 
\end{center}

\end{itemize}

\section{Discussion}

\subsection{Comparison with previous outbursts}

Be/X-ray binaries may show two types of X-ray outbursts which, in the generally accepted terminology, are known as Type I and Type
II (after Stella, White \& Rosner 1986). Sometimes they are referred to as
`normal' and `giant' (after Motch et al. 1991). Type I outbursts are
tightly linked to the orbital phase as they occur at or close to
periastron passage. The X-ray luminosity is well below
$10^{37}$ erg s$^{-1}$. In contrast, Type II outbursts may take place at
any orbital phase, are accompanied by strong spin up of the neutron star
and the X-ray luminosity may reach Eddington values.

The X-ray outburst discussed here is classifiable as a Type II
event. The classification as a giant outburst becomes apparent when a
comparison with previous outbursts is drawn. A 0535+26 has shown four
other such events in its 30 years of history: 
\begin{enumerate}
\item 1975 April, which led to
its discovery by the {\em Ariel V} satellite (Rosenberg et al. 1975),
\item 1980 October, when the 110-day periodicity was first suggested (Nagase et
al. 1982),
\item 1989 April, which seems to be the most intense outburst with
a peak intensity of 8.9 Crab in the energy range 2--26 keV (Sunyaev et al.
1989) and the first report on a cyclotron line was made
(Kendziorra et al. 1994), and
\item 1994 February, when a QPO was first
reported, indicative of the presence of an accretion disk (Finger et al.
1996). 
\end{enumerate}
In all these outbursts the X-ray flux exceeded the flux of the
Crab, the duration was $\sim$50 days and they occurred at random
binary phases.

To determine the absolute flux from A 0535+26 during the May/June 
2005 event the
RXTE/ASM count rates may be used. 1 ASM count s$^{-1}$ = $3 \times 10^{-10}$
erg cm$^{-2}$ s$^{-1}$ and the peak detected by ASM was $\sim$100 count
s$^{-1}$ ($\sim$1.3 Crab). Thus, assuming a distance of 2 kpc (Steele et
al. 1998) this implies a peak source luminosity of $1.3 \times 10^{37}$ erg
s$^{-1}$. The initial report of this outburst from the SWIFT observatory
(Tueller et al. 2005) stated that the source had a brightness that
exceeded 3 times the Crab in the BAT instrument energy range (15 -- 195 keV),
somewhat brighter than the ASM (1.3 -- 12.1 keV) data, suggesting a hard spectrum. A hard
spectrum has also been reported in other outbursts (Sunyaev et al. 1989,
Motch et al. 1991). We can estimate how much larger the hard
flux is expected to be with respect to the soft flux in Crab units by
using the spectral parameters derived from model fits to  observations
obtained during bright states. The spectrum of A 0535+26 during a Type II
outbursts is well represented by a a power-law component with
$\Gamma=0.8-1.1$ below $\sim$20 keV which decays exponentially above that
energy (Ricketts et al. 1975, Finger at al. 1994, Kendziorra et al. 1994).
Performing an order of magnitude calculation we consider a
power law with $\Gamma=1.0\pm0.2$, an exponential cut off with $E_{\rm
cut}=25\pm5$ keV and assume that the spectrum of the Crab is represented
by a single power law with $\Gamma=2.1$. Simulating the spectra of the
Crab and A 0535+26 using XSPEC we find that the flux in the 20 -- 100 keV range is
expected to be 2 -- 4 times larger than the flux in the range 1 -- 12 keV (in
Crab units), in agreement with the observed flux from SWIFT and ASM.

The May/June 2005 event began on TJD 3496 (6 May 2005 --
see Figure 3). This corresponds to a phase of $\sim$0.91 using the
ephemeris given in Section 4 above and hence is $\sim$10 days earlier
than the phase for a Type I outburst. Though, of course, the duration
of the outburst extends through phase 0.0 to phase $\sim$0.56 and so
encompasses the expected periastron passage. The duration of the
outburst was $72\pm5$ days, which is longer than typical for Type II events --
Type I outbursts last typically less than 15 days.  We find that this outburst
has characteristics of both Type I and Type II events.

In addition to the 110-day periodicity, another clock seems to be
present in A0536+26. As mentioned above, A0536+26 has so far shown
five major outbursts. The recurrence time is about 5 years. However,
they appear in groups of two with the third one being missing. There
were such outbursts in 1975 and 1980 but not 5 years later in 1985;
and then two more outbursts in 1989 and 1994 but not in
1999/2000. Maybe the latest 2005 outburst will be repeated 5 years
hence (around 2010) but not in 2015.

The profile of the outburst is unusual. Instead of the rapid rise and
slower decay typical of these types of outburst, in the May/June 2005 outburst
A 0535+26 shows an exponential flux rise and an even more rapid decay in flux.
One possibility is that the slower rise is due to
the formation of an accretion disk around the neutron star. Conversely, the 1994 Type II outburst was preceeded by a series of
smaller outbursts and the resultant rise in X-ray emission was very
much more rapid. So the accretion disk in some form may already have been
present by the time the giant ourburst began.

\subsection{Circumstellar disk}

Following on the work in Haigh et al. (2004) it is possible to use the
H$\alpha$ peak size and separation to estimate the circumstellar disk,
or at least the part emitting in the H$\alpha$ line.
The mass of the companion was assumed to be 20\mbox{$M_{\odot}$}; then
supposing a mass function of 1.64 (Finger et al. 1996) an inclination
for the orbit of $27.0\,^{\circ}$ is derived.  (Note that we have used
different values for the mass of the Be star and the inclination of
the orbit to those used by Haigh et al. (2004).)

Assuming a Keplerian velocity distribution of the matter in the
circumstellar disk, we can use the peak separation of the
doubly-peaked H$\alpha$ emission lines to gauge the radius of the
H$\alpha$ emitting region (Huang, 1972).  Then $\Delta V=2v_{obs}$
where $v_{obs}$ is the projected velocity at the outer edge of the
H$\alpha$ emitting region.  The true velocity is then $v_d =
\frac{v_{obs}}{sin i}$ where $i$, the inclination of the disk, is
assumed to be the same as the inclination of the orbit:
$27.0\,^{\circ}$.  The radius of the outer edge of the H$\alpha$
emitting region of the disk is then given by:

\begin{equation}
r=\frac{GM_{*}sin^{2}i}{{(0.5\Delta V)}^2} 
\end{equation}   

It is known that a correlation between peak separation ($\Delta V$)
and the equivalent width of the H$\alpha$ emission line exists
(Hanuschik 1989, Zamanov et al. 2001).  We used a linear relationship
between these emission line properties to derive values for the peak
separation, and hence the radius, for observations where the data were
of a quality to allow a measurement of the H$\alpha$ equivalent width
but not of the H$\alpha$ $\Delta V$.

It is important to note that there is strong evidence that the
circumstellar disk temperature remains largely unchanged throughout
most of the period of time covered by this work. This is evidenced by the
(J-K) data presented in the lower panel of Figure~\ref{fig:ir} which
only show a significant deviation from (J-K)$\sim$-0.5 during the time
of the major disk loss.

\begin{figure}\begin{center}
\includegraphics[width=70mm,angle=90]{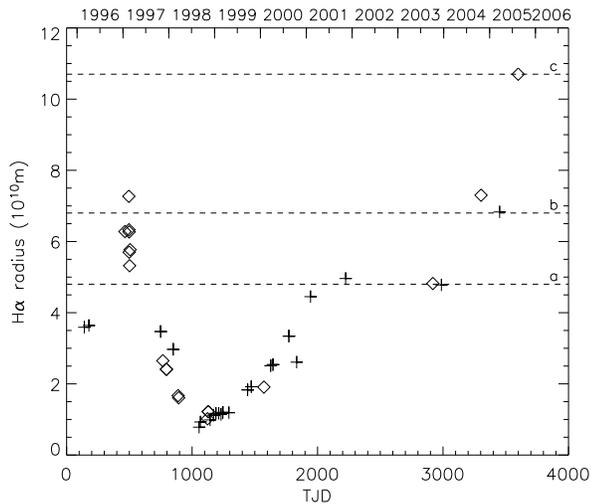}
\caption{Estimate of the radius of the H$\alpha$ emitting region of
the the disk.
Plus symbols indicate where direct measurements of $\Delta V$ were
used to determine the radius; diamond symbols indicate where the
method described in the text was employed to estimate the radius.
Dotted lines correspond to the radii present in
Figure~\ref{fig:orbit}} 
\label{fig:nick} 
\end{center} 
\end{figure}

Figure~\ref{fig:nick} shows the H$\alpha$ emitting radius as a
function of time.  Plus symbols indicate where direct measurements of
$\Delta V$ were used to determine the radius while diamonds indicate
where the method described above was employed to estimate the radius.
Values of radius are consistently below those in Haigh et al. (2004)
due to the different orbital parameters used in this paper.

\begin{figure}\begin{center}
\includegraphics[width=75mm,angle=-0]{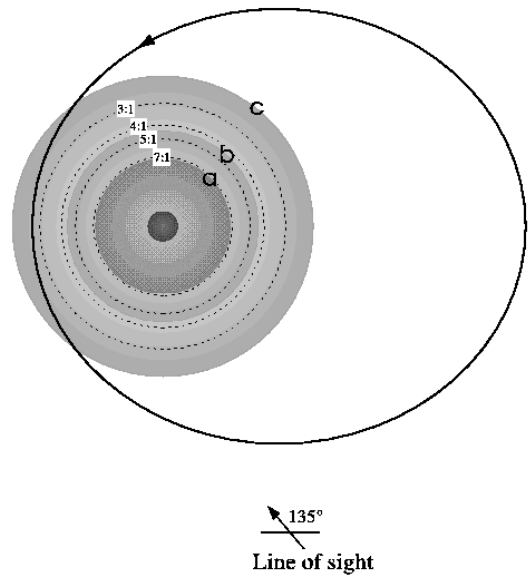}

\caption{Estimate of the radius of the H$\alpha$ emitting region of
the circumstellar disk compared to the neutron star orbit (solid black
line). The three sizes of disk shown correspond to the following radii
values : (a) $4.8\times10^{10}$m, (b) $6.8\times10^{10}$m and 
(c) $10.7\times10^{10}$m. The dashed circles are 4 truncation radii:
from inside out they are 7:1, 5:1, 4:1 and 3:1.}

\label{fig:orbit}
\end{center}
\end{figure}

Figure~\ref{fig:orbit} shows the relative sizes of the H$\alpha$
emitting regions disk (deduced from H$\alpha$ measurements) during
three different epochs, the largest (disk $c$) being the size in
August 2005, shortly after the outburst.  The middle disk (disk $b$)
is the size in March 2005, shortly before the outburst, while the
smaller disk (disk $a$) is the size during the period of December 2001
to December 2003 (no measurements are available between this last date
and March 2005). The orbital shape comes from the solution in Finger
et al. (1996).

The H$\alpha$ emitting disk size indicated by $a$ is at the 7:1
resonance radius of Okazaki and Negueruela's (2001) model of A 0535+26.
The disk remained stable at this radius for at least 2 years (see
Figure~\ref{fig:nick}) prior to the May/June 2005 outburst.  The
truncation radius places a limit on the absolute size of the
circumstellar disk, while measuring $\Delta V$ gives us an indication
of the H$\alpha$ emitting region.  If these two radii are the same, as
they seem to be in the case of A 0535+26, it suggests that the entire
disk emits H$\alpha$ radiation.

For accretion of matter onto the neutron star to occur in sufficient
amounts to cause a Type I outburst, the circumstellar disk must reach
the 4:1 resonance radius (Okazaki and Negueruela, 2001). Disk $b$,
measured $\sim$45 days before the outburst, is half way between the
5:1 and 4:1 radii, and probably growing; given that the May/June
outburst (using the ephemeris presented in this work from the RXTE/ASM
data) started at phase 0.91 (peaking at phase 0.17 and ending at phase
0.56), it seems probable that the disk had grown to at least the 4:1
resonance radius. Furthermore, the fact that the outburst was
comparable in brightness to previous Type II outbursts exhibited by
this system (1.3\,Crab compared to $\sim$1.5 - 4.5\,Crab), and also
longer (72 vs $\sim$55\,days) suggests that this is a peculiar Type II
outbursts, as the disk continued to grow as the outburst progressed.
This could explain the continued rise in X-ray luminosity through the
first $\sim$40 days of the outburst and its subsequent rapid deline
when the neutron star exited the circumstellar disk with only residual matter
in its own accretion disk to fuel the X-ray emission.

The mechanism for the May/June 2005 outburst appears to be different
to that which caused the 1994 outburst (as described by Haigh et
al. 2004). In 1994 the disk went from one truncation radius to the
next smallest, thus allowing the matter between these two truncation
radii to be accreted by the neutron star.  There is no evidence for
this occuring in the May/June 2005 outburst.

\subsection{Orbital modulation during X-ray quiescence}

The study of Be/X-ray binaries in quiescence is still an unexplored field.
So far only a handful of BeX systems have been observed in quiescence:
A0538-66, 4U 0115+63, V0332+53 (Campana et al. 2002), A0535+262
(Negueruela et al. 2000), GRO J2058+42 (Wilson et al. 2005) and SAX
J2103.5+4545 (Reig et al. 2005).

According to the standard model for Be/X-ray binaries the
circumstellar disk around the Be star's equator provides the reservoir
of matter that ultimately is accreted onto the neutron star and
converted into X-rays.  Thus, one would expect that during the X-ray
quiescent state the disk would be missing or largely debilitated. In
either case the propeller effect (Illarionov \& Sunyaev 1975) should
prevent the matter from being accreted. However, the propeller
mechanism does not seem to be at work in A 0535+26, probably due to
its relatively long spin period. Previous works by Negueruela et
al. (2000) and Orlandini et al. (2004) have demonstrated that it is
possible to detect this source during quiescent states. While X-ray
pulsations during this low-intensity states have been reported (Motch
et al, 1991), this is the first time that the quiescent emission is
shown to be modulated with the orbital period of the system.

The theoretical explanation for quiescent X-ray emission has been put
forward by Ikhsanov (2001), who shows that the accreting plasma can enter
the magnetosphere through magnetic line reconnection. Under this condition
the quiescent X-ray luminosity of A 0535+26 can be explained provided the
mass capture rate by the neutron star from the wind of the Be companion is
$\sim$$10^{-9}$ $\msun$ yr$^{-1}$. This condition is definitely satisfied
at orbital phase 0, i.e., at periastron, while although not impossible it
is more difficult to achieve at other orbital phases.

It is interesting to compare our values for the ephemeris of
periastron determined from the low-level RXTE X-ray modulation with
that determined by Finger et al. (1996). The latter authors used three
Type I outbursts detected by BATSE to determine the values for
the times of periastron (see their Table 1). Extrapolating their
values over $\sim$10 years to July/Aug 2005 results in a phase
discrepancy between the ephemeris presented here, and their prediction
for periastron, of $\delta$t = 33 $\pm$ 13d. Note that the width the peak
representing periastron passage in Figure~\ref{fig:jose} is 0.2 in phase
($\sim$22d), and hence had we chosen one of the extreme edges of the peak
then this discrepancy would only be $\delta$t = 21 $\pm$ 13d.

\section{Conclusions}

The results presented here show the recovery of strong X-ray emission
from the A 0535+26 system after 10 years of quiescence. Although it is
known that A 0535+26 exhibits X-ray activity in this state, this is  the
first time an orbital modulation has been detected in the quiescent X-ray
(using data from the RXTE/ASM). The period found at $110.0\pm0.5$ days is in
agreement with that calculated by Finger et al. (1996) from a series of
Type I outbursts in 1994. The May/June outburst began $\sim$10 days before
periastron; however, it is classified as a Type II outburst in this work
due to its high luminosity ($1.3\times10^{37}$ erg s$^{-1}$) and long duration
($\sim$72 days, the longest this system has undergone); its hard spectrum is
also consistent with previous Type II outbursts. Optical data taken shortly
after outburst show clear evidence that the Be star's circumstellar disk is
still very large, extending beyond the point of periastron passage. 


\section{Acknowledgements}

This paper uses observations made from the South African Astronomical
Observatory (SAAO), and quick-look results provided by the ASM/RXTE team.
VAM acknowledges support from the South African NRF and the British Council
in the form of a SALT/Stobie studentship. Skinakas Observatory is a
collaborative project of the University of Crete, the Foundation for Research
and Technology-Hellas and the Max-Planck-Institut f\"ur Extraterrestrische
Physik.

\bsp

\label{lastpage}

\end{document}